\documentclass[amsmath,twocolumn,superscriptaddress,showpacs,showkeys,reprint,prb,floatfix]{revtex4-1}
\usepackage{graphicx}
\usepackage{color}
\usepackage{verbatim}
\usepackage{subfigure}
\usepackage{mathbbol}
\usepackage[utf8]{inputenc}
\usepackage{float}

\usepackage{hyperref}
\newcommand{\bra}[1]{\ensuremath{\left\langle #1\right|}}
\newcommand{\ket}[1]{\ensuremath{\left|#1\right\rangle}}

\newlength\fheight 
    \newlength\fwidth 
    
\begin{document}
\title{Stochastic Bloch-Redfield theory: quantum jumps in a solid-state environment}
\author{Nicolas \surname{Vogt}} 
\affiliation{Institut f\"ur Theorie der Kondensierten Materie,  \\ Karlsruhe Institute of Technology, D-76128 Karlsruhe, Germany}
\affiliation{DFG-Center for Functional Nanostructures (CFN), Karlsruhe Institute of Technology, D-76128 Karlsruhe, Germany}
\author{Jan Jeske}
\affiliation{Chemical and Quantum Physics, School of Applied Sciences, \\ RMIT University, Melbourne, Australia}
\author{Jared H. \surname{Cole}}
\affiliation{Chemical and Quantum Physics, School of Applied Sciences, \\ RMIT University, Melbourne, Australia}
\begin{abstract}
	We discuss mapping the Bloch-Redfield master-equation to Lindblad form and then unravelling the resulting evolution into a stochastic Schrödinger equation according to the quantum-jump method. We give two approximations under which this mapping is valid.  This approach enables us to study solid-state-systems of much larger sizes than is possible with the standard Bloch-Redfield master-equation, while still providing a systematic method for obtaining the jump operators and corresponding rates. We also show how the stochastic unravelling of the Bloch-Redfield equations becomes the kinetic Monte Carlo (KMC) algorithm in the secular approximation when the system-bath-coupling operators are given by tunnelling-operators between system-eigenstates. 
	The stochastic unravelling is compared to the conventional Bloch-Redfield approach with the superconducting single electron transistor (SSET) as an example.
\end{abstract}

\date{September 20, 2013}
\maketitle
 \setlength\fheight{0.65\columnwidth} 
    \setlength\fwidth{0.8\columnwidth} 
\section{Introduction}
In almost all experimentally accessible few-state quantum systems: atoms in optical cavities, qubits in superconducting circuits, quantum dots, Rydberg atoms and many other systems, the quantum system is in contact with a large environmental bath. The interaction with the bath leads to a loss of phase-coherence between the quantum states and to relaxation. The time-evolution of an open quantum system given by the master-equation for the density matrix is much more complex than the time-evolution of a closed quantum system of the same size $N$. In particular in solid-state-physics, the nature of the bath and the coupling between system and bath can be very complicated\cite{Makhlin2004,Shnirman2005}. In that case it is not sufficient to use a phenomenological master-equation of Lindblad-form\cite{Lindblad1976}. With the Bloch-Redfield equation\cite{Bloch1957,Redfield1957} a powerful tool has been developed to obtain the master-equation from the microscopic parameters of the model in a unified way. 

In the last twenty years a rich theory has been developed, especially in quantum optics, to unravel the differential $N\times N$ matrix-master-equation in Lindblad form\cite{Diosi1985,Diosi1986,Dalibard1992,Teich1992,Wiseman1993a,Brun2002,Gambetta2005} or in the  non-Markovian-case\cite{Gambetta2002,Gambetta2003,Gambetta2004} into a stochastic Schrödinger equation (SSE) for a state vector of size $N$. The time-evolution of the density-matrix can be obtained from the unravelling by averaging over many stochastic realisations of the time-evolution of the system (called trajectories) given by the SSE. In measurement theory\cite{Wiseman1993} where the bath is (partially) given by a measurement device the stochastic unravelling is used to treat the interaction with the constantly measured environment as a succession of stochastic events which depend on the previous trajectory. In general the unravelling can be used to gain a numerical advantage over the standard master-equation for large systems. Calculating each trajectory scales with the system size as $\mathcal{O}(N^2)$, solving the master equation as $\mathcal{O}(N^4)$. 

Stochastic unravellings have been considered in solid-state-systems\cite{Korotkov1999,Goan2001, Korotkov2001a, Korotkov2001, Oxtoby2003, Korotkov2003, Oxtoby2005, Oxtoby2006, Oxtoby2008}, mostly in the context of measurement theory,  either by obtaining the parameters of a Lindblad-master-equation for the specific microscopic bath model or by assuming a phenomenological Lindblad-master-equation. 
In the first part of this paper, sections \ref{sec:QJ} and \ref{sec:BR} we discuss the general unravelling of the generic Bloch-Redfield equation into the form of a stochastic Schrödinger equation. This stochastic Bloch-Redfield approach is general in the sense that it defines an algorithm to obtain a valid stochastic unravelling from the same parameters of the microscopic model that enter the Bloch-Redfield master-equation. In the second part Sec.\ref{sec:SSET} we compare the stochastic Bloch-Redfield algorithm with the Bloch-Redfield master-equation with the example of a superconducting single electron transistor (SSET)\cite{Brink1991,Schon1994} and demonstrate that the stochastic Bloch-Redfield approach is able to handle much larger system sizes than the master-equation approach.

\section{Quantum jumps in the Lindblad equation}
\label{sec:QJ}
The density matrix of a physical system always has the form $\rho(t) = \sum_{\phi} p_{\phi} \ket{\phi}\bra{\phi}$ .
The most general time independent master equation that conserves this form for a physical initial density matrix $\rho_0$ is given by the Lindblad equation\cite{Lindblad1976}:
\begin{eqnarray}
	\label{eqn:Lindblad}
	\dot{\rho} &=& i \left[ \rho, H_S\right] + \sum_{\alpha} \Gamma_{\alpha} \left( L_{\alpha}\rho L^{\dagger}_{\alpha}  - \frac{1}{2}\left\{L^{\dagger}_{\alpha}L_{\alpha},\rho\right\} \right)  \ , \nonumber \\
\end{eqnarray}
where the Lindblad operators $L_{\alpha}$ and the rates $\Gamma_{\alpha}$ determine the decoherence properties of the system. The Lindblad operators and rates are at the level of the Lindblad equation free parameters of the system model, they either have to be introduced as phenomenological variables from physical insight or obtained from a microscopic model of the bath by other means. The Lindblad equation itself does not  contain a way to obtain these parameters from a microscopic model.
In many cases for example in most quantum optics models where the bath is the quantized light field the form of the possible Lindblad operators arises naturally from the coupling between the open quantum system and the environment. 

The Lindblad master equation can be rewritten as a differential vector-equation 
\begin{eqnarray}
	\dot{\vec{\rho}} &=& \mathcal{L}\vec{\rho} \ ,
\end{eqnarray}
where $\mathcal{L}$ is the Lindblad superoperator, for an $N$-dimensional open quantum system an $N^2 \times N^2$-matrix. The computational complexity of the Lindblad master-equation therefore scales as $\mathcal{O}(N^4)$ with the system size, making it impossible to numerically solve the master-equation of open quantum system with a somewhat larger Hilbert space.

To overcome this problem stochastic unravellings of the Lindblad master equation have been developed\cite{Diosi1985,Diosi1986,Dalibard1992,Percival1998,Brun2002,Gambetta2005}. Instead of calculating the time-evolution of the density matrix $\rho(t)$ a stochastic trajectory of a system state $\ket{\psi(t)}$ is calculated from a stochastic Schrödinger equation (SSE). The density matrix is obtained by averaging over the pure state density matrices of the states from many stochastic trajectories
\begin{eqnarray}
	\rho(t) &=& \frac{1}{m} \sum \ket{\psi(t)}\bra{\psi(t)} \ ,
\end{eqnarray}
where $m$ is the number of stochastic trajectories. A single state $\ket{\psi(t)}$ is an $N$-element vector and the SSE has computational complexity $\mathcal{O}(N^2)$ and  the stochastic unravelling has a total numerical complexity of $\mathcal{O}(m N^2)$. The stochastic unravelling scales much better with the size of the Hilbert space than the standard Lindblad master equation and it becomes possible  to study larger open quantum system which are not accessible to the density-matrix method.

The stochastic unravelling of the master equation \eqref{eqn:Lindblad} is not unique. There are unravellings based on continuous stochastic time evolution like quantum state diffusion\cite{Percival1998} as well as unravellings which rely on deterministic time-evolution between stochastic events like the quantum jump method\cite{Dalibard1992,Brun2002,Gambetta2005}. Here we briefly introduce the quantum jump algorithm on which the stochastic Bloch-Redfield method introduced later on in this paper is based. Extensive discussion of the quantum jump method can be found in references [\onlinecite{Brun2002}] and [\onlinecite{Wiseman2009}].

In the quantum jump approach  a system with $n_{\alpha}$ Lindblad operators in a state $\ket{\psi(t)}$ can over the infinitesimal time $dt$ evolve into $n_{\alpha}+1$ states $\ket{\psi(t+dt)}^0$ and $\ket{\psi(t+dt)}^{\alpha}$:
\begin{eqnarray}
	\label{eqn:FullLinddtStates}
	|\psi(t+dt)\rangle^{0} &=& \frac{1}{\sqrt{p_0}} (\mathbb{1}-i H_{co}dt) |\psi(t)\rangle \label{eqn:psi0} \\
	H_{co} &=& H + i \frac{1}{2} \sum_{\alpha} \Gamma_{\alpha} L_{\alpha}^{\dagger} L_{\alpha} \\
	p_0 &=& \langle \psi(t)| \left| \mathbb{1}-i H_{co}dt\right|^2 |\psi(t)\rangle \\
	|\psi(t+dt)\rangle^{\alpha} &=& \frac{1}{\sqrt{p_\alpha}} \sqrt{\Gamma_{\alpha} dt}  L_{\alpha} |\psi(t)\rangle \nonumber \\
	&=& \frac{1}{\left|L_{\alpha}|\psi(t)\rangle\right|} L_{\alpha} |\psi(t)\rangle \\
	p_{\alpha} &=& \langle \psi(t)| \Gamma_{\alpha} L_{\alpha}^{\dagger} L_{\alpha} |\psi(t)\rangle dt \ .\label{eqn:p alpha}
\end{eqnarray}
 The time-evolution to state $\ket{\psi(t+dt)}^0$ corresponds to the part of the Lindblad equation which describes purely coherent evolution.
The state  $\ket{\psi(t+dt)}^0$ is determined by the complex Hamiltonian $H_{co}$ which combines the coherent time-evolution of the system Hamiltonian $H_S$ with the decoherence part of the Lindblad equation \eqref{eqn:Lindblad} where the Lindblad operators only act on the density matrix from one side $\left\{L^{\dagger}_{\alpha}L_{\alpha},\rho\right\} $. This combination of Lindblad operators does not take a pure state density matrix out of the pure-state subset of the set of density matrices. 

The time-evolution to one of the states $\ket{\psi(t+dt)}^{\alpha}$ is called a quantum jump. In a single trajectory it corresponds to the transition from one pure quantum state to another however after averaging over many trajectories these jumps lead to the decay of the density matrix $\rho$ from a pure quantum state to a classical mixture of quantum states. 

The probabilities $p_0$ and $p_{\alpha}$  of the state $\ket{\psi(t)}$ evolving into $\ket{\psi(t+dt)}^0$ or $\ket{\psi(t+dt)}^{\alpha}$ are given by the normalization factors
in  $\ket{\psi(t+dt)}^0$ and $\ket{\psi(t+dt)}^{\alpha}$.
The density-matrix at time $t+dt$ after averaging is given by
\begin{eqnarray}
	\rho(t+dt) &=& p_0 \rho^0(t+dt) + p_{\alpha} \rho^{\alpha}(t+dt)
\end{eqnarray}
where $\rho^0(t+dt)$ and $\rho^{\alpha}(t+dt)$ are the density matrices corresponding to the states $\ket{\psi(t+dt)}^0$ and $\ket{\psi(t+dt)}^{\alpha}$. The probability weights $p_0$ and $p_{\alpha}$ exactly compensate the normalization factors of the stochastic states and one obtains the same time-evolved density matrix as from the  Lindblad equation \eqref{eqn:Lindblad}.

Solving the non-linear differential equation \eqref{eqn:psi0} to \eqref{eqn:p alpha} and creating a random number at each time-step to decide whether a quantum jump occurs is numerically expensive.  A common variation\cite{Wiseman2009} of the quantum jump algorithm makes use of the close connection between the normalization in Eq.\ref{eqn:psi0} and the probability $p_0$. The probability that no quantum jump occurs between $t_0=0$ and time $t$ follows the same exponential decay as the quadratic state norm of the state $\ket{\psi_{u.n.}(t)}$ that evolves according to Eq.\ref{eqn:psi0} without the normalization.
\begin{eqnarray}
	\label{eqn:ProbandNormTimeDep}
	P_0(0,t) &\propto& \exp\left[ \ln(p_0) \frac{t}{\Delta t} \right]  \\
	\left| |\psi_{u.n.}(t)\rangle \right|^2 &\propto& \exp\left[- 2 \ln\left( \frac{1}{\sqrt{p_0}}\right)+\ln\left(\frac{t}{\Delta t}\right) \right] \nonumber \\
	& & = \exp\left[ \ln(p_0) \frac{t}{\Delta t} \right] \ .
\end{eqnarray}
The time evolution in Eq.\ref{eqn:psi0} is linearized so that the state-norm decays over time, once the norm decreases below the value of a random number $r_1 \in \left(0, 1\right)$ at $t_1$ the time evolution is stopped, $\ket{\psi(t_1)}$ normalized and it is decided with the help of a second random number $r_2 \in \left(0, 1\right)$ and the probabilities $p_{\alpha}$ which quantum jump occurs. All states are normalized after the full time-evolution has been calculated.

\section{Stochastic Bloch-Redfield equations}
\label{sec:BR}
In solid state physics it is often necessary to derive the master equation of an open quantum system from the microscopic properties of the environment and the coupling between the environment and the system. In contrast to quantum optical systems, the form of the decoherence part of the master equation can not easily be guessed when the environment is complicated\cite{Makhlin2004,Paladino2002,Shnirman2005,Paladino2007} and may strongly depend on time-varying parameters of the system \cite{Vogt2012}. The Bloch-Redfield equation\cite{Bloch1957,Redfield1957} is a master equation for an open quantum system that directly incorporates the properties of the microscopic model. 

The Bloch-Redfield approach starts with a world Hamiltonian $H_W$ consisting of system ($H_S$), bath ($H_B$) and coupling ($H_{SB}$) Hamiltonian,
\begin{eqnarray}
	H_W &=& H_S+H_B+H_{SB} \label{eqn:WorldHam} \\
	H_{SB} &=& \sum_{i} g_i X_i z_i \label{eqn:CouplingHam} \ .
\end{eqnarray}
The coupling Hamiltonian is a linear combination of products of system coupling operators $z_i$ operating on the open system and bath coupling operators $X_i$ with the coupling strengths $g_i$. Note that the coupling operators themselves need not be Hermitian as long as the full coupling Hamiltonian is.

Assuming that the system bath coupling strength is a small parameter compared to the other parameters of the system the von-Neuman equation for the time evolution of the world density matrix $\dot{\rho}_W(t) = i \left[H_W, \rho_W(t) \right]$ in the interaction picture is expanded into an integro-differential equation in second order of the coupling Hamiltonian. Assuming that the world density matrix is always a tensor product of a time-dependent system density matrix and a constant bath density matrix $\rho_W(t) = \rho_S(t) \otimes \rho_B$ the environmental degrees of freedom are traced out to obtain a equation of motion for the system density matrix. A detailed derivation can be found in [\onlinecite{Carmichael1999}].
\begin{eqnarray}
	\dot{\rho}_S(t) &=& i \left[\rho_S(t),H_S\right] \nonumber \\ && - \int_{0}^{\infty} d\tau \sum_{ij} \tilde{C}_{ij}(-\tau) Z_{ij}(t,\tau) -\tilde{C}_{ji}(\tau) \bar{Z}_{ij}(t,\tau)  \label{eqn:BR_Schr_tau_int} \label{eqn:BR_tau} \nonumber \\ 
\end{eqnarray}
with
\begin{eqnarray}
	Z_{ij}(t,\tau) &=& \rho_S(t) z_i e^{i H_S \tau} z_je^{-i H_S \tau} - z_i \rho_S(t)e^{i H_S \tau}z_je^{-i H_S \tau} \nonumber \\
	\bar{Z}_{ij}(t,\tau) &=&  e^{H_S \tau}z_j e^{-i H_S \tau}\rho_S(t)z_i - e^{i H_S \tau} z_je^{-iH_S \tau}z_i \rho_S(t)  \nonumber 
\end{eqnarray}
and the bath operator correlation function
\begin{eqnarray}
	\tilde{C}_{ij}(t,t') &=& \tilde{C}_{ij}(t-t') = \textrm{Tr}\left[ X_i(t-t') X_j(0) \rho_B\right] \label{eqn:corr_fun}
\end{eqnarray}
In the time-integral form of the Bloch-Redfield equation \eqref{eqn:BR_tau} the Markov approximation in the interaction picture was used. In this approximation it is assumed that the correlation functions Eq.\ref{eqn:corr_fun} decay on a much shorter timescale than the timescale of relaxation $T_1$ and dephasing $T_2$ of the open quantum system and the timescale of the coherent evolution due to the coupling Hamiltonian $H_{SB}$. From here on we also assume that the correlation function $\tilde{C}_{ij}(\tau)$ is only nonzero if the i-th and j-th coupling operators are Hermitian conjugates $z_i = z_j^{\dagger}$. This is no limitation in practice as in most physical cases the coupling operators are either observables and themselves Hermitian operators or are of Jaynes-Cummings type $\left(\sigma^+a+\sigma^-a^{\dagger}\right)$ for which the assumption also holds.   

Usually the time-integral form of the Bloch-Redfield equation is rewritten with the help of the spectral function $C_{ij}(\omega)$, the eigenenergy differences of the system $\omega_{\beta \gamma}$ and the system coupling operators in the eigenbasis of the system 
$\xi_i=V^\dagger z_i V$, where $V$ diagonalizes the system Hamiltonian $V^\dagger H_S V = \text{diag}(E_1, E_2, \dots)$.
The states $\ket{\beta}$, $\ket{\gamma}$, $\ket{\delta}$ and $\ket{\eta}$ always refer to eigenstates of the open quantum system $H_S$.
\begin{eqnarray}
	C_{ij}(\omega) &=& \int_{-\infty}^{\infty}  \tilde{C}_{ij}(\tau) e^{-i \omega \tau} d\tau \label{eqn:SpecFunDef} \\
		\omega_{\beta \gamma} &=& E_{\gamma}-E_{\beta} \nonumber \\
	\xi_i^{\beta \gamma} &=& \langle \beta | \xi_i | \gamma \rangle \nonumber \ .
\end{eqnarray}
Introducing an infinitesimal real element into the time-dependent exponential of Eq.\ref{eqn:BR_tau} the time-integral can be evaluated and the Bloch-Redfield equation takes the form Eq.\ref{eqn:BR_omega} where an imaginary contribution due to a Cauchy principal value has been ignored since it only introduces a slight Lamb-shift effect in the system. 
\begin{eqnarray}
	\dot{\rho}_S(t) &=& i \left[\rho_S(t),H_S\right] 
	-\frac{1}{2} \sum_{i,j} \sum_{\beta, \gamma, \delta, \eta} \xi_j^{\beta,\gamma} \xi_i^{\eta,\delta}  \nonumber \\ &&\left\{ C_{ij}(-\omega_{\beta \gamma}) D_{\beta \gamma \delta \eta}(t) - C_{ji}(\omega_{\beta \gamma})\bar{D}_{\beta \gamma \delta \eta}(t)  \right\} \ , \nonumber \\ \label{eqn:BR_omega} 
	D_{\beta \gamma \delta \eta}(t) &=& \rho_S(t)  |\eta\rangle\langle\delta| |\beta\rangle\langle\gamma| - |\eta\rangle\langle\delta| \rho_S(t) |\beta\rangle\langle\gamma|  \nonumber \\
	\bar{D}_{\beta \gamma \delta \eta}(t) &=& |\beta\rangle\langle\gamma| \rho_S(t) |\eta\rangle\langle\delta| - |\beta\rangle\langle\gamma| |\eta\rangle\langle\delta|\rho_S(t) 
\end{eqnarray}

The Bloch-Redfield equation can be simplified further by applying the secular approximation. In the secular approximation combinations of transitions $\ket{\beta}\bra{\gamma}$ and $\ket{\eta}\bra{\delta}$ in the coupling operators $\xi_i$ and $\xi_j$ in Eq.\ref{eqn:BR_omega} are suppressed  when the associated eigenenergy differences $\omega_{\beta \gamma}$ and $\omega_{\eta \delta}$ do not cancel each other. Mathematically a Kronecker delta is introduced in the sum over the eigenstates in Eq.\ref{eqn:BR_omega}:  
\begin{eqnarray}
	\sum_{\beta, \gamma, \delta, \eta} &\rightarrow& \sum_{\beta, \gamma, \delta, \eta} \delta(\omega_{\beta \gamma}-\omega_{\delta \eta}) \label{eqn:sec_appr_strict} \ .
\end{eqnarray}
The physical justification for this approximation is the occurrence of an oscillating phase $\exp\left( i (\omega_{\beta \gamma}- \omega_{\delta \eta})t \right)$ in the Bloch-Redfield equation in the interaction picture. As long as $\omega_{\beta \gamma}-\omega_{\delta \eta}$ is large compared to the decoherence rates and coherent frequencies in the interaction picture the contribution of the corresponding terms in the Bloch-Redfield equation averages to zero due to the fast oscillations of the terms. 
This also means that the secular approximation should not be applied to terms where $\omega_{\beta \gamma}$ and $\omega_{\delta \eta}$ only differ by a small amount. However for the moment we will understand by secular approximation the strict version of Eq.\ref{eqn:sec_appr_strict}.

%We now proceed to unravel the Bloch-Redfield equation \ref{eqn:BR_omega} into a stochastic Schrödinger equation.

We now proceed to discuss stochastic unravellings of master-equations in a structured environment such as described by the Bloch-Redfield equation \ref{eqn:BR_omega}.
 On a general note we point out that stochastic unravellings on non-Markovian master equations have already been introduced using either a quantum state diffusion method\cite{Diosi1997,Diosi1998} or additional degrees of freedom coupled to the system\cite{Imamoglu1994}. As the Bloch-Redfield equation with the structured environment given by the spectral function $C_{ij}(\omega)$ is just a special case of a non-Markovian master equation these unravellings are in principle able to handle the microscopic details required in solid-state master equations. However since these methods are designed to handle non-Markovian problems they are computationally much more expansive, requiring time integrals at each time-step or expanding the Hilbert space of the system. Therefore in this work we will focus on a systematic way to unravel a specifically Markovian master-equation of a structured environment. 

The general Bloch-Redfield equation can produce unphysical density-matrices under adverse circumstances  and can not be rewritten in Lindblad form~\cite{Whitney2008}. It is thus not generally possible to unravel the Bloch-Redfield equation without further approximations (or additional computational complexity).  Fundamentally, this is because 
the density matrices constructed by averaging over many SSE state-trajectories always have the form of physical density matrices $\rho = \sum_{m} \frac{1}{m} \ket{\psi_m}\bra{\psi_m}$ as they are by construction a statistical mixture of pure state density matrices $\ket{\psi_m}\bra{\psi_m}$. As the Lindblad equation is the most general equation that produces density matrices of this form we rewrite the master equation in Lindblad form to find a valid stochastic unravelling. 

We now present two approximations which let us recast the Bloch-Redfield equation in Lindblad form.  First we discuss the stochastic unravelling in the secular approximation for which it is well known\cite{Carmichael1999, Jeske2013} that the Bloch-Redfield equation can be rewritten in Lindblad form. Thereafter we discuss the piecewise flat spectral-function (PWFS) approximation which also leads to a Lindblad form and is less strict than the secular approximation. Furthermore in the PWFS approximation the number of Lindblad operators in the rewritten master equation does not scale as $\mathcal{O}(N^2)$ with the system size as in the secular approximation which can be important in practical numerical applications. It is important to note here that although the resulting quantum jump equations derive from a Lindblad form equation, the rates and jump-operators are not phenomenological parameters but are derived in a systematic way from the microscopic properties of the model in the same fashion as for the Bloch-Redfield master equation.

To apply the secular approximation in a systematic way we must more carefully define which states are considered (quasi-)degenerate and which are not. We define subsets $\mathcal{M}(\omega)$ of $N\times N$-matrices for each eigenenergy-difference $\omega$ of the open quantum system:
\begin{eqnarray}
	\mathcal{M}(\omega) &=& \left\{ M \in \mathbb{C}^{N \times N} \ | \ \forall \ \beta, \gamma \ \ \omega_{\beta \gamma} \ne \omega \ \Rightarrow \ M_{\beta \gamma} = 0 \right\} \, \nonumber \\ \label{eqn:Qjumps:Decomposition}
\end{eqnarray}
and the corresponding projectors into each subset $\mathcal{P}(\omega)$. To apply the secular approximation we  decompose each original coupling operator $\xi_i$ into $m_E$ coupling operators, where $m_E$ is the number of unique eigenenergy differences $\omega$ of the system. Each of the new coupling operators $\xi_k^J$ is an element of one of the subsets $\mathcal{M}(\omega)$. In the new decomposition the correlation functions can be rewritten as the rate matrix $\tilde{\Gamma}_{k l}$ that also contains the Kronecker delta of the secular approximation. 
\begin{eqnarray}
	\xi_k^J &=& \xi_{\left(m_E (i-1)+n\right)}^J = \mathcal{P}(\omega_n) \xi_i  \label{eqn:Coupling_operator_decomposistion} \\
	\tilde{\Gamma}_{k l} &=& C_{i(k,l) j(k,l)}\left(- \omega(k) \right) \delta(\omega_{\beta \gamma}-\omega_{\delta \eta})
\end{eqnarray}

The secular approximation simply requires that $\tilde{\Gamma}_{k l}$ is only non-zero for all combinations of coupling operators $\xi^J_k$ and $\xi^J_l$ from conjugate subsets $\mathcal{M}(\omega(k))$ and $\mathcal{M}(\omega(l))=\mathcal{M}(-\omega(k))$. From this and the assumption that the correlation function  $C_{ij}$ is only non-zero for Hermitian conjugate operators $\xi_{i}$ and $\xi_j$ it follows that for all nonzero $\tilde{\Gamma}_{k l}$ we have $\xi_{k}^J = \left(\xi_l^J\right)^{\dagger} $. Since $\tilde{\Gamma}_{kl}$ can only be non-zero for one combination of $k$ and $l$ we can write $\Gamma_{k} = \tilde{\Gamma}_{kl}$ and the Bloch-Redfield equation takes the form
\begin{eqnarray}
	\dot{\rho}_{S}(t) &=& i \left[\rho_{S}(t),H_S\right]  +\frac{1}{2} \sum_{k}  \Gamma_{k} \left[ 2 \xi_k^J \rho_S(t) (\xi_{k}^{J})^\dagger \right. \nonumber \\ & &\left. \quad - \rho_S(t)  (\xi_k^{J})^{\dagger} \xi_k^J -  \xi_k^J  (\xi_{k}^{J})^\dagger \rho_S(t) \right]  \ . \label{eqn:BR_qjump} 
\end{eqnarray}
The Bloch-Redfield equation has become a Lindblad equation and we can apply the usual quantum jump algorithm. The new coupling operators $\xi_k^J$ are the jump-operators of the quantum jump algorithm.

Here we consider independent (uncorrelated) noise contributions but a brief comment on the more general case [\onlinecite{Jeske2013}] is warranted. In this case the previous assumption that the correlation function $\tilde{C}_{ij}(\tau)$ is only non-zero if $z_i=z_j^\dagger$ must be dropped. However in the secular approximation one then finds that those new coupling operators which are projected onto opposite frequencies $\xi_k^J=\mathcal{P}(\omega)\xi_i$ and $\xi_l^J=\mathcal{P}(-\omega)\xi_i$ are Hermitian conjugates of each other $\xi_k^J=(\xi_l^J)^\dagger$. This can be for two reasons; either the corresponding original coupling operator is Hermitian $\xi_i=\xi_i^\dagger$ and therefore:
\begin{align*}
[\mathcal{P}(\omega)\xi_i]^\dagger&=\sum_{\omega_{\beta \gamma}=\omega} (\ket{\beta}\bra{\beta} \xi_i \ket{\gamma}\bra{\gamma})^\dagger \\
&= \sum_{\omega_{\beta \gamma}=\omega}\ket{\gamma}\bra{\gamma} \xi_i \ket{\beta}\bra{\beta} = \mathcal{P}(-\omega)\xi_i
\end{align*}
or, as in the case of Jaynes-Cummings type coupling operators, the two coupling operators which are already Hermitian conjugates are each entirely projected into subsets of opposite frequency. With spatially correlated decoherence, Eq.\ref{eqn:BR_qjump} then sums over two indices:
\begin{eqnarray}
	\dot{\rho}_{S}(t) &=& i \left[\rho_{S}(t),H_S\right]  +\frac{1}{2} \sum_{k,l}  \tilde{\Gamma}_{kl} \left[ 2 \xi_k^J \rho_S(t) (\xi_{l}^{J})^\dagger \right. \nonumber \\ & &\left. \quad - \rho_S(t)  (\xi_l^{J})^{\dagger} \xi_k^J -  \xi_k^J  (\xi_{l}^{J})^\dagger \rho_S(t) \right]  \ . \label{eqn:BR_qjump_spat_cor} 
\end{eqnarray}
This can be brought back into the form of Eq.\ref{eqn:BR_qjump} by diagonalisation of the coefficient matrix $u^\dagger \tilde{\Gamma}_{kl} u=\text{diag}( \Gamma_1,  \Gamma_2, \dots)$ with the unitary matrix $u$. Eq.\ref{eqn:BR_qjump} is then recovered by replacing the coupling operators $\xi_k^J \rightarrow \tilde{\xi}_k^J = \sum_l u_{lk} \xi^J_l$. (see ref.~\onlinecite{Jeske2013}). With this extra step, any form of spatially correlated decoherence can be evaluated as a stochastic B-R equation. However, the numerical effort of the diagonalisation of $\tilde{\Gamma}_{kl}$ depends on the degree of spatial correlation and on $m_E$. In many systems the eigenenergy differences can be grouped such that $m_E=3$, leading to a maximal numerical scaling of $\mathcal{O}(N^2)$, however the worst case would be $m_E=N^2$, leading to $\mathcal{O}(N^4)$.

Having rewritten the Bloch-Redfield equation in the form Eq.\ref{eqn:BR_qjump} it would also be possible to use a quantum state diffusion unraveling\cite{Percival1998} to obtain a stochastic Schrödinger equation. However this unravelling leads to a non-linear differential equation with a stochastic component in each time-step which is computationally more expensive than a quantum-jump unravelling and in most cases the results are equivalent.

%Herein lies the key advantage of a stochastic B-R approach.  
The key advantage of a stochastic Bloch-Redfield approach lies in the following: 
One need only start with the form of the system-environment coupling operator and the bath noise correlation function and then systematically derive all jump operators and their associated rates.  This procedure is unchanged for arbitrary numbers of components within the system and is especially suited to solid-state systems with many states over an extended region of space.

So far the stochastic unravelling of the Bloch-Redfield equation relied on the strict secular approximation, which is problematic when a system has two or more eigenenergy differences $\omega_{\beta \gamma}$ and $\omega_{\delta \eta}$ which are not equal but whose frequency difference $\Delta_{\beta \gamma \delta \eta} = \omega_{\beta \gamma}-\omega_{\delta \eta} $   corresponds to an oscillation on timescales equal to or larger than the timescale of the time-evolution of the system in the interaction picture.  These cases are especially likely to occur in systems with a large Hilbert space for which the stochastic unravelling is designed where diagonalization can lead to a comparably dense but not degenerate spectrum. An example is a tight binding chain with a thousand sites compared to a model with only five sites. In both cases the energy band has the same width but in the first case the band contains a much larger number of eigenenergies.

It is not (always) necessary to use the full strict secular approximation to rewrite the Bloch-Redfield equation in Lindblad form. For the derivation of the Bloch-Redfield equation to be valid the correlation function $\tilde{C}_{ij}(\tau)$ needs to decay fast on the timescale of time-evolution of the density matrix in the interaction picture. It follows that the Fourier transformed,  spectral function needs to be smooth on the frequency-scales of the coherent time evolution in the interaction picture and the maximal relaxation $\Gamma_1$ and dephasing $\Gamma_2$ rates of the open quantum system. To keep this approximation consistent with our earlier discussion on the secular approximation we must quantify the concept of ``smoothness''. Consequently we assume we can find a piecewise flat approximation of the spectral function
\begin{eqnarray}
	C_{ij}(\omega) &\approx& C^{pwf}_{ij}(\omega) =   \sum_n C_{ij}^n \mathcal{B}_n(\omega) \ , \label{eqn:CorrelationFunctionBinApproximation}  \\
	\mathcal{B}_n(\omega)  &=& \sum_i \Theta(\omega-\omega_{n,i}^{<}) \Theta(\omega_{n,i}^{>}-\omega) 
\end{eqnarray}
so that the width of each bin $\omega_{n,i}^{>}-\omega_{n,i}^{<}$ is larger than $\Gamma_1$, $\Gamma_2$ and the frequencies of the coherent time-evolution in the interaction picture. The function $\mathcal{B}_n(\omega) $ defines the family of all bins corresponding to the value $C_{ij}^n$ in the approximated spectral function. An example of this approximation is shown in Fig.\ref{fig:spec_fun_approx}.

We now apply what we call the piecewise flat spectral function (PWFS) approximation and neglect all combination of transitions $\ket{\beta}\bra{\gamma}$ and $\ket{\eta}\bra{\delta}$ in the Bloch-Redfield equation whose transition frequency and negative transition frequency  $\omega_{\beta \gamma}$ and $-\omega_{\delta \eta}$ belong to different bins in Eq.\ref{eqn:CorrelationFunctionBinApproximation}. For all transitions from bins that are not next neighbours and the vast majority of transitions from neighbouring bins this is justified as  $\omega_{\beta \gamma}-\omega_{\delta \eta}$ is large and the secular approximation is valid. The only problematic case arises when $\omega_{\beta \gamma}$ and $-\omega_{\delta \eta}$ are both close to a bin boundary $\omega_{n,i}^{<}$. We assume that it is either possible to chose the bin boundaries in such a way that there are no transition frequencies close to the boundary on at least one side of the boundary or that, in the case where this is not possible and the distribution of transition frequencies is dense, the fraction of pairs for which the PWFS approximation is not justified is so small compared to the number of all transitions that the error introduced in the time-evolution by this approximation is negligible. In practice this assumption holds for most systems.

We can now use the same type of decomposition of the coupling operators as we did in the secular approximation, however the subsets of the matrix-space are now defined by the rates $\Gamma$ corresponding to the amplitude of the piecewise flat spectral function in one family of bins $\mathcal{B}_n(\omega) $:
\begin{eqnarray}
	\mathcal{M}(\Gamma) &=& \left\{ M \in \mathbb{C}^{N\times N} \ | \ C^{pwf}_{ij}(\omega_{\beta \gamma}) \ne \Gamma \Rightarrow  M_{\beta \gamma} = 0 \right\} \nonumber \\
\end{eqnarray}
We define the projector $\mathcal{P}(\Gamma)$ and the jump-operators $\xi_k$ in the same way as before and obtain the Lindblad equation \eqref{eqn:BR_qjump}.

\section{Link to kinetic Monte Carlo}
In this section we show that  in the limit of quantum jumps being limited to transitions between eigenstates, stochastic Bloch-Redfield reduces to the well known kinetic Monte Carlo method.
The kinetic Monte Carlo (KMC) algorithm has for a long time been used very successfully in solid state physics\cite{Bakhvalov1989,wasshuber2001} and other branches of physics and chemistry\cite{Voter2005,Reuter2011} to simulate systems whose time-evolution is determined by incoherent tunnelling processes between metastable states. The KMC method can not deal with the coherent time-evolution of the system as it deals only with incoherent transitions between a given set of basis states.

The kinetic Monte Carlo algorithm simulates a system of $N$ states by selecting the state of the system at time $t+\Delta t$ and the transition time $\Delta t$ from a probability distribution based on the transition rates from the state at time $t$ to all other states of the system. Let $\vec{s}$ be the vector of the sums over the tunnelling rates form state $j$ to states $l$,
\begin{eqnarray}
	s_k &=& \sum_{l=1}^{k} \Gamma_{jl}  
\end{eqnarray}
The next state $k$ is chosen with the random number $r_1 \in \left(0,1\right)$ so that
\begin{eqnarray}
	s_k &\le& s_N \cdot r_1 < s_{k+1} \ .
\end{eqnarray}
The probability to tunnel into state $k$ is proportional to the tunnelling rate $\Gamma_{jk}$.
The probability of the system leaving state $j$ by tunneling into any other state after the escape time $\Delta t$ is given by
\begin{eqnarray}
	p_{esc}(\Delta t) &=& s_N e^{-s_N \Delta t} \label{eqn:KMC_escape_time_distribution} \ .
\end{eqnarray}
Instead of drawing $\Delta t$ from an exponential distribution, the time that is attributed to the tunneling process can be chosen with the evenly distributed random number $r_2 \in \left(0,1\right)$, such that:
\begin{eqnarray}
	\Delta t &=& - s_N \cdot \ln(r_2)  \label{eqn:KMC_escape_time}
\end{eqnarray}
Comparing equations Eq.\ref{eqn:KMC_escape_time_distribution} and Eq.\ref{eqn:ProbandNormTimeDep} we see that the kinetic Monte Carlo algorithm and the quantum jump algorithm differ only in two points. In the KMC algorithm the order of the choice of jump state and jump time is inverted which is purely a question of convention. Secondly in the KMC algorithm the jump time can be calculated with the tunneling rates which are known immediately after the previous tunnelling process and are constant for $\Delta t$. Contrary to the quantum jump algorithm the time-evolution between tunnelling processes does not have to be computed in the KMC-method.

To show the equivalence we start with the strict secular approximation and we now only consider systems where the jump operators $\xi_k^J$ from Eq.\ref{eqn:Coupling_operator_decomposistion} correspond to exactly one transition between eigenstates:
\begin{eqnarray}
	\xi_k^J &\propto& |\beta\rangle \langle\gamma| \label{eqn:KMC_jump_operator_restriction} \ ,
\end{eqnarray}
This is for example the case if the system has no degenerate transition frequencies and from $\omega_{\beta \gamma}= \omega_{\eta \delta}$ follows $\ket{\beta} = \ket{\eta}$ and $\ket{\gamma} = \ket{\delta}$. It is guaranteed that the system after each quantum jump at time $t_i$ is in an eigenstate $\ket{\beta_i}$ with the complex phase $\phi_i$. The time-dependence of the state between two quantum jumps is then given by: 
\begin{eqnarray}
	\forall \  t \in \left[t_i, t_{i+1}\right] \quad|\psi(t)\rangle &=& e^{i \phi_i} e^{-E_{\beta_i}(t-t_i)} |\beta_i\rangle \label{eqn:Qjumps:KMC_state_time_evolution} \ .
\end{eqnarray}
The differential equation for the unnormalized system state $\ket{\psi_{u.n.}(t)}$ is easily solved analytically and one obtains an analytic expression for the time of the next quantum jump $t_{i+1}$ as a function of the random number $r_i \in \left(0,1\right)$ and the complex phase $\phi_{i+1}$ at $t_{i+1}$.
\begin{eqnarray}
	\frac{d}{dt}|\psi_{u.n.}(t)\rangle &=& -i E_{\beta_i} - \frac{1}{2} \sum_{k} \Gamma_k \left(\xi_k^J\right)^{\dagger} \xi_k^J |\psi_{u.n.}(t)\rangle \nonumber \\ \\
	|| | \psi_{u.n.}(t) \rangle||^2 &=& \exp \left( - \sum_{k} \Gamma_k \langle \beta_i | \left(\xi_k^J\right)^{\dagger} \xi_k^J | \beta_i \rangle (t-t_i) \right) \nonumber \\ \\
	t_{i+1} &=& t_i+\ln(\frac{1}{r_i}) \left( \sum_k \Gamma_k \langle \beta_i|\left(\xi_k^J\right)^\dagger \xi_k^J|\beta_i\rangle  \right)^{-1} \label{eqn:Qjumps:KMC_jump_time} \\
	\phi_{i+1} &=& \phi_i-E_{\beta_i}(t_{i+1}-t_i)
\end{eqnarray}
It is no longer necessary to solve a complicated differential equation of complexity $\mathcal{O}(N^2)$ to obtain the time-evolution between quantum jumps. The quantum-jump algorithm simplifies to a KMC-algorithm where we simply have to additionally keep track of the phase $\phi_i$ and the energy $E_{\beta_i}$ between KMC-steps. A  similar approach that relies on finding a time dependent basis that diagonalizes the density matrix at all points in time has been discussed by Teich and Mahler\cite{Teich1992}.

It is important to note here that this does not mean one can easily turn every KMC-simulation into a stochastic Bloch-Redfield simulation and obtain information about the coherent time-evolution of the system without additional numerical cost. The stochastic Bloch-Redfield method always requires the diagonalization of the system Hamiltonian $H_S$ to obtain the transition frequencies $\omega_{\beta \gamma}$ and the coupling operators $\xi_j$. This places an upper bound on the system size $N$ although it is much larger than the upper bound for the direct solution of the density matrix master equation. In a KMC simulation this is not necessary as long as the original basis is a good approximation of the eigenbasis, in fact is it not even necessary to keep track of all basis states, at any one point in time one only needs the current state and all states connected to this state by non-zero tunnelling rates. It is for example possible to simulate  quasi-particle transport through a fifty-site array with KMC\cite{Bylander2005, Bylander2006}. A system with a Hilbert space size of $2^{50}$ is far beyond the scope of the stochastic Bloch-Redfield method.

\section{Numerical example: The SSET}
\label{sec:SSET}
The superconducting single electron transistor (SSET) is a mesoscopic superconducting device consisting of two superconducting leads connected to a superconducting island with two Josephson junctions with capacitances $C_J$ and Josephson energy $E_j$. The superconducting island is connected to a ground capacitance $C_g$ which is connected to the ground by the voltage source $V_g$. The left and right leads are biased at voltage $V_1$ and $V_2$ respectively.

The SSET has been studied extensively in the last twenty five years\cite{Brink1991,Schon1994,Nakamura1997,Choi2003}. The fact that their behaviour is well known, that they are a solid-state system with a structured bath and that they can have arbitrary large Hilbert spaces when all tunnelled charge states are treated coherently makes them an ideal test object for the stochastic Bloch-Redfield method.

The state of an SSET can be given by the number of charges on the superconducting island $N$ and the number or elementary charges that have tunnelled through the right Josephson junction $\bar{N}$. We limit the Hilbert space to the states with only one Cooper pair, anti Cooper pair, quasi-particle or anti quasi-particle on the island, $N = -2 \dots 2$. The size of the Hilbert space is $5 m_{\bar{N}}$ where $m_{\bar{N}}$ is the number of tunnelled-charge-states considered.
In terms of $N$ and $\bar{N}$ the system Hamiltonian consisting of the island charging $H_c$, the voltage $H_V$ and the tunneling $H_t$ Hamiltonian is given by
\begin{eqnarray}
	H_S &=& H_c+H_V+H_t \label{eqn:SSET_H} \\
	H_c &=& \frac{1}{2 C} \left(e \hat{N} - n_g\right)^2  \label{eqn:Single_Dot_charge_H}  \\  
	H_V &=&  V_2  e \hat{\bar{N}} - V_1 \left( e \hat{\bar{N}} - e \hat{N}\right) \label{eqn:Single_Dot_Voltage_H}  \\
	H_t &=& \sum_{N,\bar{N}} \left( E_J \ket{N+2}\bra{N} \ket{\bar{N}-2}\bra{\bar{N}} \right. \nonumber \\ 
	& &\left.+ E_J \ket{N+2}\bra{N} +  \textrm{h.c.} \right) \label{eqn:Single_Dot_tunneling_H} \\
	n_g &=&  C_g V_g + C_J (V_2-V_1) \label{eqn:Single_Dot_offset_charge} \\
	C &=& 2 C_J+C_g  \label{eqn:Single_Dot_Capacitance} \ ,
\end{eqnarray} 
where $n_g$ is the offset charge on the island and $C$ is the total capacitance of the superconducting dot.
\subsection{The Josephson Quasi-Particle Cycle}
For the first set of simulations we consider transport through the SSET in the Josephson quasi-particle (JQP) cycle\cite{Brink1991,Schon1994}. The voltage on the left lead is set to zero $V_1 = 0$ and $V_g$ is chosen so that the offset charge is $n_g=1$ and of all states with the same number of tunnelled charges $\bar{N}$ the state with one quasi-particle on the island $\ket{N=1,\bar{N}}$ has the lowest energy. The right lead is biased with the voltage $V_2 = E_c+2 \Delta$ where $E_c = \frac{e^2}{2 C}$ is the charging energy of the island and $\Delta$ is the superconducting gap. 

Transport in the JQP cycle is a two stage process. The states $\ket{N=0,\bar{N}}$ and $\ket{N=2,\bar{N}}$ are degenerate and the tunneling Hamiltonian $H_t$ causes coherent oscillations of Cooper-pairs across the left lead. Coherent tunnelling across the right lead is largely suppressed because of the large energy difference between states $\ket{N=0,\bar{N}}$ and $\ket{N=2, \bar{N}\pm 2}$. From state $\ket{N=2,\bar{N}}$ the system relaxes to state $\ket{N=0,\bar{N}=2}$ via two quasi-particle-tunnelling-processes and the JQP cycle can start again. 

In this system the environmental bath is given by the equilibrium quasi-particles in the leads and on the island that do no contribute  to the charging Hamiltonian $H_C$ of the SSET. The pairs of system coupling operators $(z_i , z_j)$ with non-zero correlation functions  are given by $\left( \ket{N,\bar{N}}\bra{N-1,\bar{N}+1},  \ket{N,\bar{N}}\bra{N+1,\bar{N}-1}  \right)$ for tunnelling over the right lead and  $\left( \ket{N,\bar{N}}\bra{N-1,\bar{N}},  \ket{N,\bar{N}}\bra{N+1,\bar{N}}  \right)$ for tunneling over the left lead. The corresponding spectral functions are given by\cite{Schon1994}
\begin{eqnarray}
	C(\omega) &=& \frac{1}{e^2 R_t} \int_{-\infty}^{\infty} d\epsilon \int_{-\infty}^{\infty} d\epsilon' \nonumber \\ & & \quad \mathcal{N}(\epsilon) \mathcal{N}(\epsilon') f(\epsilon) \left[ 1- f(\epsilon')\right] \delta(\epsilon-\epsilon'-\omega) \ , \label{eqn:QP_tunnelling_rate}
\end{eqnarray}
where $R_t$ is the Josephson junction tunnelling resistance, $\mathcal{N}$ is the density of states of the quasi-particles and $f(\epsilon)$ is the Fermi-distribution. Eq.\ref{eqn:QP_tunnelling_rate} can be calculated numerically, but here we are more interested in the general comparison of the stochastic Bloch-Redfield method with the master-equation solution than in the finer details of SSET physics. We therefore take the low temperature limit and approximate the correlation function
\begin{eqnarray}
	C(\omega) &=& \frac{1}{e^2 R_t} \Theta(\omega -2 \Delta) \omega \ , \label{eqn:QP_rate}
\end{eqnarray}
The superconducting gap $\Delta$ in the density of states of the quasi-particles manifests itself in the Heaviside function that suppresses quasi-tunneling unless the energy difference between the involved states allows for the breakup of one Cooper pair into two quasi-particles. Due to this suppression, no incoherent quasi-particle tunnelling processes occur over the left lead as the energies do not allow the creation of two quasi-particles. Instead of the inverse tunnelling resistance, from now on we use the parameter $\Gamma = \frac{2 \Delta}{e^2 R_t} $ to characterise the strength of the quasi-particle tunnelling.

\begin{figure}
\centering
\begin{comment}
\begin{tikzpicture}
	\node (img1) {\input{figures/jqpc_Nbar_combined_conv_paper.tikz}};
	\node (img2) at (img1.east) [xshift=-2.1cm,yshift=-0.9cm]  {\resizebox{0.4\columnwidth}{!}{\input{figures/SSET.tikz}} };
\end{tikzpicture}
\end{comment}
\includegraphics{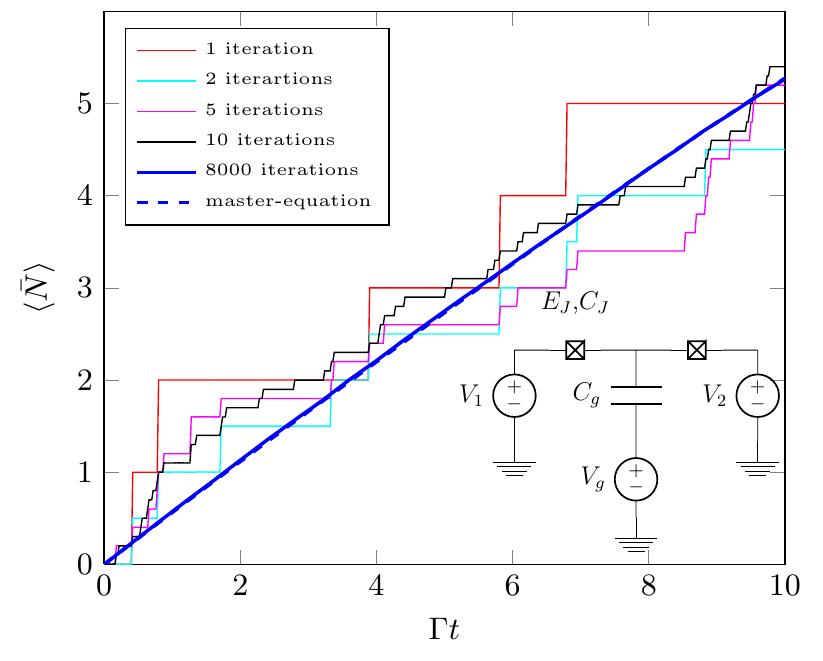}
\caption{ \label{fig:jqpc_Nbar_exp} A plot of the convergence of the expectation value of the time-evolution of the tunnelled charges $\left\langle \bar{N} \right\rangle$ in the JQP cycle when averaged over an increasing number of iterations of the stochastic Bloch-Redfield algorithm in the PWFS approximation. After $8000$ iterations the result is almost indistinguishable from the solution of the master-equation (dashed blue line).  The expectation value $\left\langle \bar{N} \right\rangle$, which corresponds to the transported charge, increase linearly over time. \\
Inlay: Circuit diagram of the SSET with capacitance to the ground $C_g$, the bias voltages $V_1$,$V_2$, the voltage to the ground $V_g$ and the Josephson junctions with the Josephson energy $E_J$ and the capacitance $C_J$. }
\end{figure}

To compare the stochastic Bloch-Redfield method with the standard Bloch-Redfield master-equation we simulated a system with $\bar{N} = 0\dots10$ and $55$ basis states with the Bloch-Redfield equation and a standard ODE-solver. We compare this to the stochastic Bloch-Redfield algorithm in the secular approximation and in the PWFS approximation. In terms of the Josephson energy, parameters were chosen as $E_c = 2.5 E_j$, $\Gamma = 0.2 E_j$ and $\Delta = 20 E_j$. Both stochastic Bloch-Redfield algorithms were averaged over 8000 trajectories. In figure Fig.\ref{fig:jqpc_Nbar_exp}  we show that the expectation value of $\bar{N}$ changes in discrete jumps, corresponding to the quantum jumps, in a single trajectory. One can also see how the time-dependence of the expectation value converges to the correct result over several iterations.

\begin{figure}
\centering

	\includegraphics{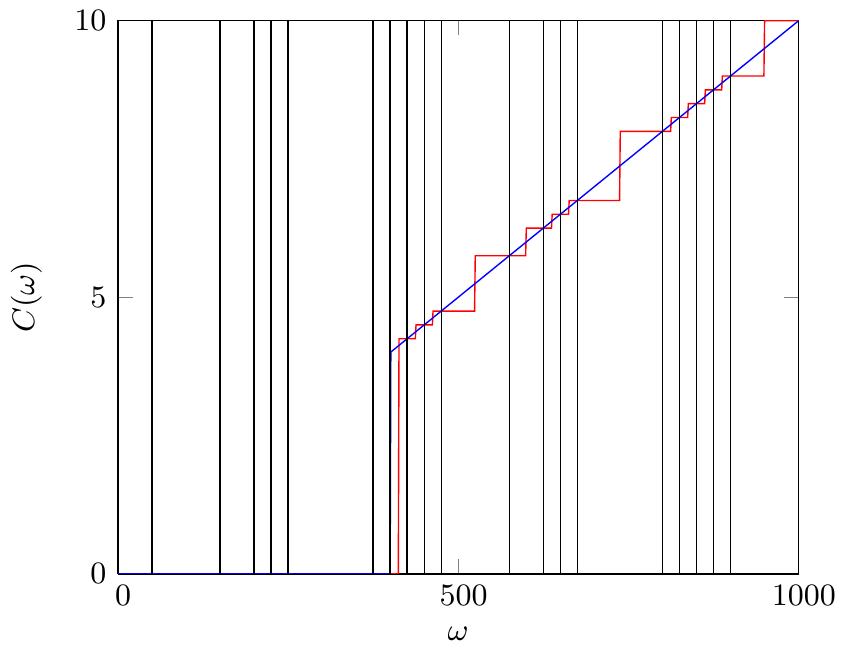}

	\caption{\label{fig:spec_fun_approx} The piecewise flat spectral function approximation: The spectral function $C(\omega)$ (blue) is approximated by the piecewise-flat function $C^{pwf}(\omega)$ (red) defined by the values of $C(\omega)$ at the transition frequencies $\omega_{\beta \gamma}$ of the system (black lines). The transition frequencies in the JQPC-example are spaced so far apart that each transition frequency corresponds to another bin in $C^{pwf}(\omega)$. In the intervals between the transition frequencies $C^{pwf}(\omega)$ can take very different values from $C(\omega)$, however this is not important as the value of the spectral function at these energies does not enter the Bloch-Redfield equation.}
\end{figure}

\begin{figure}
\centering
\begin{comment}
\subfigure[\label{fig:jqpc_N_single_run_paper}]{

\input{figures/jqpc_N_single_run_paper.tikz}}

\subfigure[\label{fig:jqpc_N_pop_short}]{

\input{figures/jqpc_N_pop_short.tikz}}
\end{comment}
%\begin{comment}
\subfigure[\label{fig:jqpc_N_single_run_paper}]{

\includegraphics{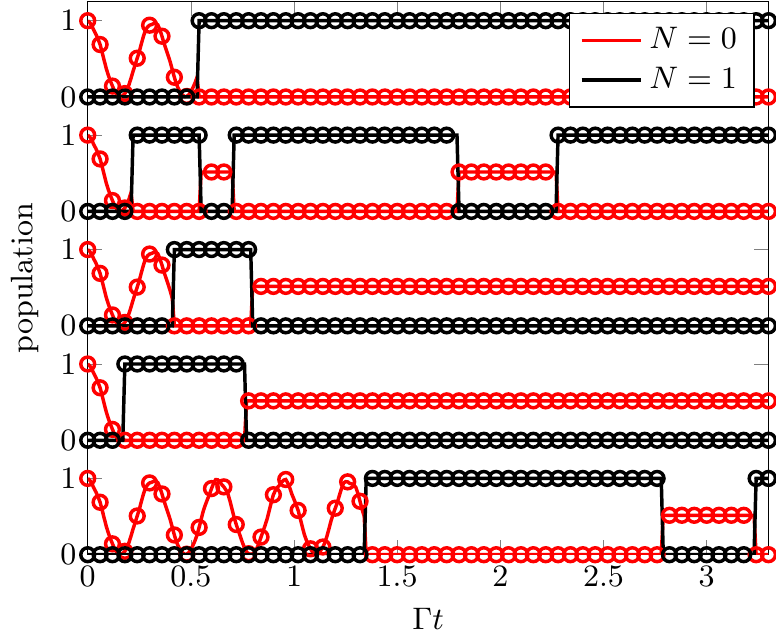}}

\subfigure[\label{fig:jqpc_N_pop_short}]{
\includegraphics{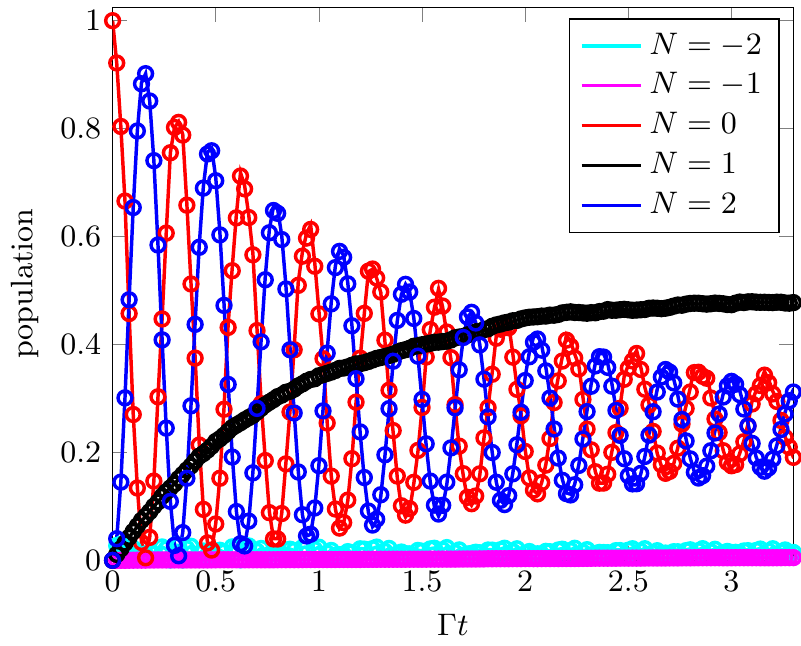}
}
%\end{comment}
\caption{\label{fig:jqpc_N_single_pop_short} $(a)$: The population in the states $N=0$ and $N=1$ of the SSET for several single trajectories of the stochastic Bloch-Redfield algorithm. The population is plotted with an offset for each trajectory. For the chosen parameter of the JQP cycle the PWFS (solid lines)  and the secular approximation  (circles)  are equivalent as explained in Fig.\ref{fig:spec_fun_approx}. The PWFS-approximation and the secular trajectories were initialized with the same seed for the random number generator. As the system was not initialized in an eigenstate the quantum jump trajectories show coherent oscillations of the system state before the system is projected into an eigenstate by the first quantum jump. \\
$(b)$: The time-evolution of the population of the charge-states of the SSET-island $N=-2$ to $N=2$ in the JQP cycle over a short time obtained by averaging over $8000$ of the trajectories seen in Fig.\ref{fig:jqpc_N_single_run_paper} (circles). The master-equation gives the same result (solid line) as the stochastic Bloch-Redfield algorithm in the PFS-approximation . Population oscillates coherently between states $\left| N=0 \right\rangle$ and $\left| N=2 \right\rangle$. The amplitude of the oscillation decays as dephasing destroys the coherent quantum-oscillations. Population relaxes from states $\left|N=0\right\rangle$ and $\left|N=2\right\rangle$ to state $\left| N=1 \right\rangle$ via dissipative quasi-particle tunnelling.  
}
\end{figure}

In the JQPC-simulations the secular approximation and the PWFS-approximation lead to the same set of jump operators. The linear dependence of spectral function Eq.\ref{eqn:QP_rate} leads to very small bin sizes in Eq.\ref{eqn:CorrelationFunctionBinApproximation} so that each bin corresponds to only one transition frequency and both approximations are equivalent (see also Fig.\ref{fig:spec_fun_approx}). As a consequence the trajectories in Fig.\ref{fig:jqpc_N_single_run_paper}  corresponding to the two approximations are exactly equal as long as the same seed for the random number generator is used for both trajectories. In both trajectories a quantum jump always projects the system into an eigenstate and no coherent oscillations appear between quantum jumps. 
After averaging the trajectories of the occupation states of the SSET island from  Fig.\ref{fig:jqpc_N_single_run_paper} lead to coherent oscillation of the population in the $\ket{N}$-states of the SSET that are damped by the interaction with the environment (compare Fig.\ref{fig:jqpc_N_pop_short}). In contrast to Fig.\ref{fig:jqpc_Nbar_exp} which only shows the increase in the number of tunnelled charges due to dissipative transport Fig.\ref{fig:jqpc_N_single_pop_short} demonstrates how quantum mechanical oscillations evolves from averaging over many quantum jump trajectories. Those trajectories themselves  show no further oscillations after the first quantum jump which projects the system into an eigenstate.

\begin{figure}[h]
\centering
\begin{comment}
\begin{tikzpicture}
	\node (img1)  {\input{figures/jqpc_Nbar_long_run_gaussian_fit.tikz}};
	\node (img2) at (img1.east) [xshift=-2.25cm,yshift=1.9cm]  {\resizebox{0.4\columnwidth}{!}{\input{figures/jqpc_Nbar_long_run_exp.tikz}}};
\end{tikzpicture}
\end{comment}
\includegraphics{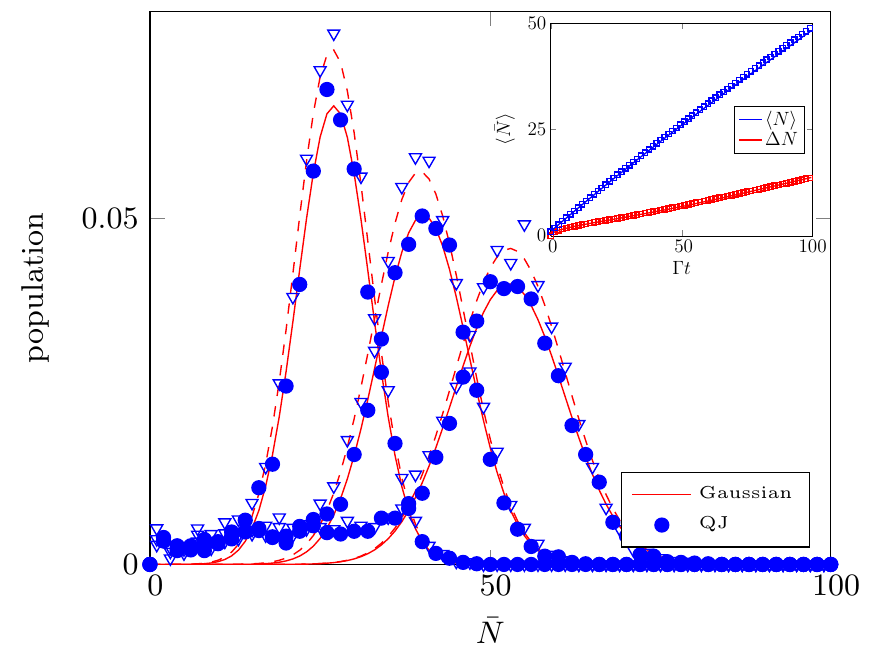}
\caption{\label{fig:jqpc_Nbar_long_run_exp}
The distribution of the population at $\Gamma t = 50$,  $\Gamma t = 75$ and  $\Gamma t = 100$ of the $\left|\bar{N} \right\rangle$-states. The distribution has a peak that moves to larger $\bar{N}$ and broadens over time as charge is transported. The form of the peak can be fitted to the sum of two Gaussians with the same width and peak position but different amplitudes for odd (dots and solid line) and even (triangles and dashed line) $\bar{N}$-states. Although we consider a slightly different quantity this shows a similarity to the behaviour of the probability distribution of $N$-charges tunnelling over a time-interval $\tau$ in an SSET that has completely equilibrated  predicted by Choi et al. [\onlinecite{Choi2003}].\\
Inlay:  Plot of the expectation value  $\left\langle \bar{N} \right\rangle$ (blue/light) and the standard deviation $\Delta \bar{N}$ (red/dark) for a large system ($\bar{N}_{max} = 100$) and long simulation time $\Gamma t_{max} = 100$ not accessible with numerical solution to the full density matrix master equation. Results of the stochastic Bloch-Redfield algorithm in the secular (squares) and PFS (circles) approximation after $8000$ iterations are shown. The expectation value $\left\langle \bar{N} \right\rangle$ and $\Delta \bar{N}$ increase linearly for the whole simulation as expected\cite{Choi2003}. }
\end{figure}

To demonstrate the power of the stochastic Bloch-Redfield methods we also used them to simulate a much larger system of 505 basis states ($\bar{N} = 0\dots 100$) over a time period five times as long as in the previous simulations. This system size is beyond the scope of standard numerical master-equation solutions. The inlay of Fig.\ref{fig:jqpc_Nbar_long_run_exp}  shows how charge is transported through the SSET as the expectation value $\left\langle \bar{N} \right\rangle$ increases and how the population is distributed over more $\bar{N}$-states over time. One advantage of the large system size and simulation times is that it is possible to see how the distribution of $\bar{N}$-states approaches a double Gaussian shape for odd and even $\bar{N}$-states  and how the standard deviation of the Gaussians increases linearly with time as shown in Fig.\ref{fig:jqpc_Nbar_long_run_exp}. A similar structure has been analytically obtained by Choi et al. [\onlinecite{Choi2003}] for the number of charges tunnelling through an SSET over time $\tau$ in the limit of an SSET that has been equilibrated for an infinite time. 

\begin{figure}[th!]
	\vspace{0.2cm}
	\begin{centering}
		\begin{comment}
		\begin{tikzpicture}
	\node (img1) {	\resizebox{0.8\columnwidth}{!}{\includegraphics{figures/jqpc_Nbar_long_run_3D.png}}};
	\node (a) at (-2.5,-2)  {$\Gamma t$};
	\node (b) at  (1.1,-2.2)  {$\bar{N}$};
	\node[rotate=90] (c) at  (-3.5,0.5) {population};	
\end{tikzpicture}
\end{comment}
\includegraphics{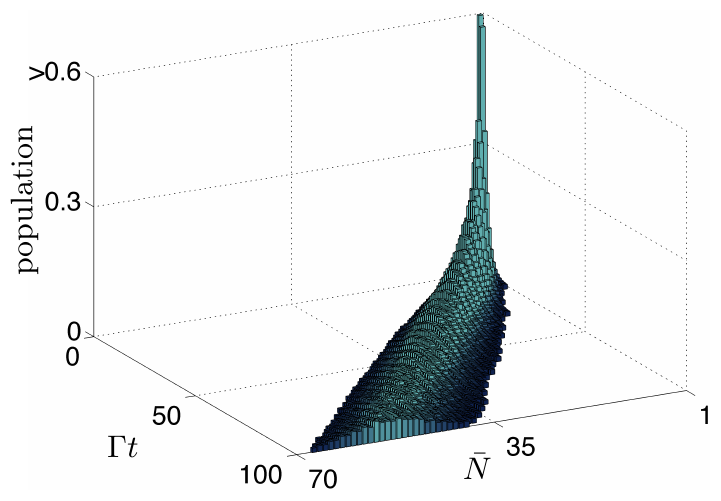}
	\caption{ \label{fig:jqpc_Nbar_long_run_3D} Three-dimensional representation of the spreading of the population (height) over the $\left|\bar{N} \right\rangle$-states (x-axis) and time (y-axis) in the stochastic Bloch-Redfield JQPC-simulation of the large system. States with population $<0.01$ have been truncated for clarity. Starting in a single peak at $\Gamma t = 0$ the population spreads out over several  $\left|\bar{N} \right\rangle$-states  over time while the center of the distribution moves to higher $\bar{N}$.}
\end{centering}
\end{figure}

\subsection{Incoherent Cooper Pair Tunnelling}

A major advantage of the Bloch-Redfield equation is that it can correctly describe the relaxation processes caused by purely longitudinal environmental noise via the diagonalization of the coupling operators $z_i$. When using the Lindblad equation or the KMC algorithm calculating the correct relaxation rates usually requires a polaron transformation and the use of $P(E)$-theory\cite{Ingold1992}. 

To demonstrate that we retain this feature in the stochastic Bloch-Redfield methods we consider the incoherent Cooper pair tunneling (ICPT) in the SSET. For simplicity we set all incoherent quasi-particle tunnelling rates to zero. We couple to the environmental noise with only one logintudinal coupling operator $z_i = N$ and use a flat spectral function for $\omega>0$ with the values for $\omega<0$ given by detailed balance
\begin{eqnarray}
	C(\omega) &=& \left\{ \begin{array}{l r}
		 \Gamma_2 & \omega \ge 0 \\
		 \Gamma_2 e^{-\beta \omega} & \omega < 0
	\end{array} \right. \\
	\beta &=& \frac{1}{k_B T}
\end{eqnarray}
Due to the coherent Cooper pair tunnelling terms in the system Hamiltonian this longitudinal coupling leads to incoherent dissipative Cooper-pair tunnelling which drives charge transport through the SSET. We chose the parameters $\Delta = 20 E_j$, $n_g = 1$, $E_c = 2.5 E_j$, $\beta = 100$, $V_1 =0$, $V_2 = 12.5 E_j$ and $\Gamma_2 = 0.2 E_j$.

\begin{figure}
\centering
\begin{comment}
\subfigure[\label{fig:icpt_N_single_pop_short_secular}]{
\input{figures/icpt_N_single_pop_short_secular.tikz}
}
\subfigure[\label{fig:icpt_N_single_pop_short_pwfs}]{
\input{figures/icpt_N_single_pop_short_pwfs.tikz}
}
\end{comment}
%\begin{comment}

\subfigure[\label{fig:icpt_N_single_pop_short_secular}]{
\includegraphics{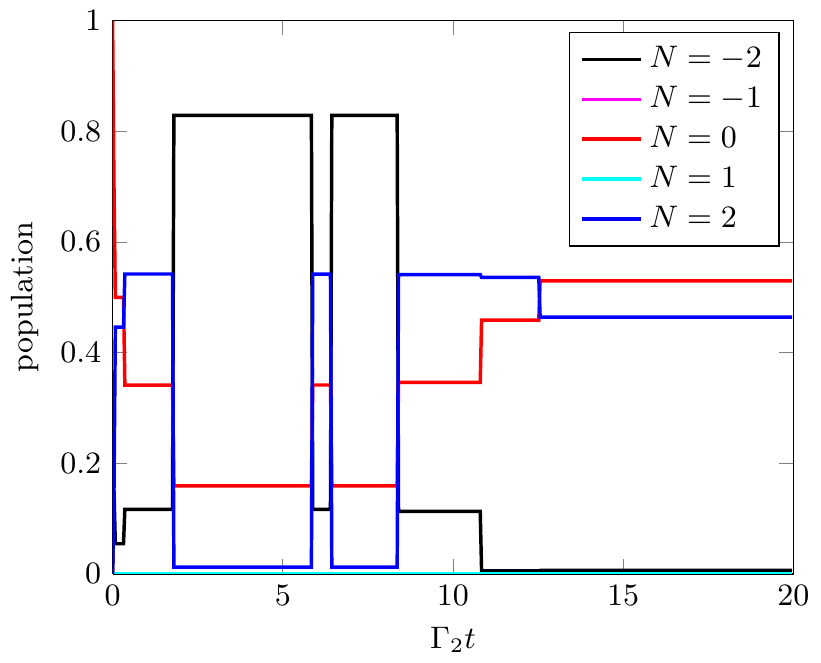}
}
\subfigure[\label{fig:icpt_N_single_pop_short_pwfs}]{
\includegraphics{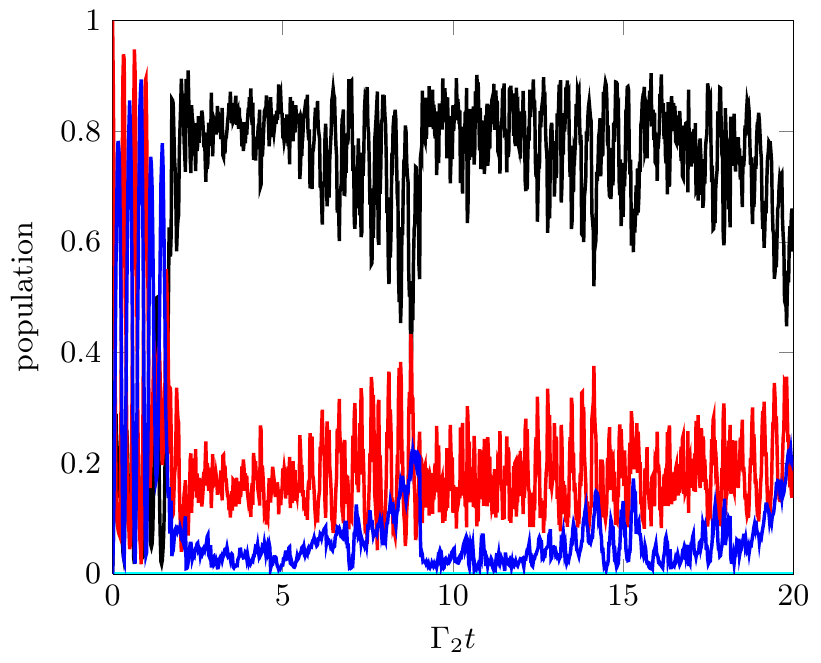}

}
%\end{comment}

\caption{\label{fig:icpt_N_single_pop_short} The stochastic trajectories of the ICPT-system in the PWFS (Fig.\ref{fig:icpt_N_single_pop_short_pwfs}) and the secular (Fig.\ref{fig:icpt_N_single_pop_short_secular}) approximation. The trajectories were initialised with different  random number seeds. The time-evolution in the secular approximation is limited to quantum jump between eigenstates whereas in the PWFS-approximation the system shows strong coherent time evolution between quantum jumps. After averaging over several thousand trajectories we obtain the same time-evolution for the density matrix as for the standard master-equation from both approximations. Both approximations are valid as the stricter secular approximation holds for the system we consider.} 
\end{figure}

In the secular approximation each quantum jump projects the system into an eigenstate as in the JQPC-simulation. Here however the PWFS-approximation is not equivalent to the secular approximation as the spectral function is already flat and $C^{pwf}(\omega)$ consists of only one bin for $\omega>0$. In the PWFS-approximation the system-state shows coherent oscillations between two quantum jumps in one stochastic trajectory, as shown in Fig.\ref{fig:icpt_N_single_pop_short}.  Averaged over several thousand trajectories one still obtains the same time-evolution of the density-matrix from the stochastic Bloch-Redfield algorithm with both approximations and the master-equation. The secular approximation holds in the considered system.

\begin{figure}[H]
	\vspace{0.2cm}
	\begin{centering}
\begin{comment}
		\begin{tikzpicture}
	\node (img1) {\resizebox{0.8\columnwidth}{!}{\includegraphics{figures/icpt_Nbar_long_run_3D.png}}};
	\node (a) at (-3,-2)  {$\Gamma t$};
	\node (b) at  (1,-3)  {$\bar{N}$};
	\node[rotate=90] (c) at  (-3.5,1) {population};	
\end{tikzpicture}
\end{comment}
\includegraphics{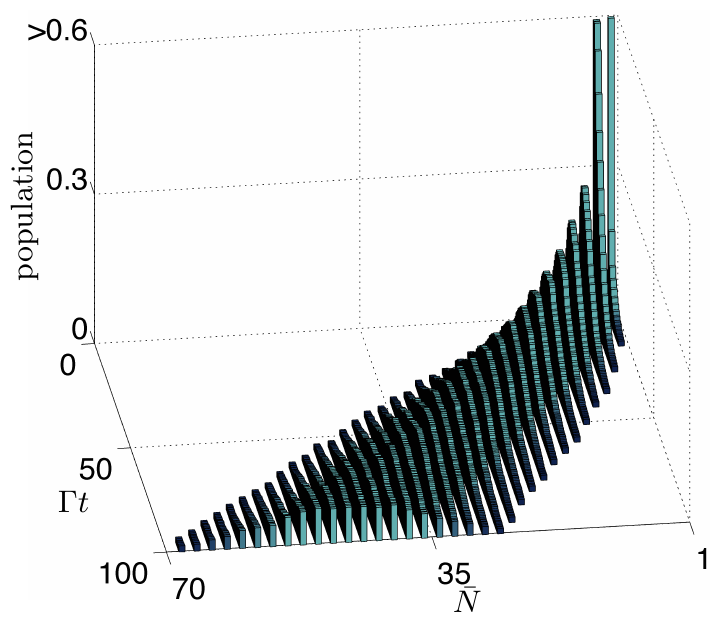}
\caption{ \label{fig:icpt_Nbar_long_run_3D} The three-dimensional representation of the population for a long simulation $\Gamma t_{max} = 500$ of a large ICPT-system $\bar{N}_{max} = 100$. The main difference between the ICPT-case and the JQPC-case Fig.\ref{fig:jqpc_Nbar_long_run_3D} is that incoherent Cooper-pair tunnelling only connects states which differ by $\bar{N} =2 $ and therefore only odd $\left|\bar{N}\right\rangle$-states are occupied in this case.}
\end{centering}
\end{figure}

Again we also show the results of a large system simulation ($\bar{N} = 0\dots100$) in Fig.\ref{fig:icpt_Nbar_long_run_3D}. As in the JQPC-case the population spreads out over the $\bar{N}$-states over time, however only odd $\bar{N}$-states are occupied. Incoherent Cooper pair tunnelling can only change $N$ and $\bar{N}$ by two electron charges and the initial state is chosen to be $\ket{N=0,\bar{N} =1}$.
\section{Conclusion}
The Bloch-Redfield master-equation is widely used in solid-state physics to model decoherence. In this work we have discussed two approximations in which it is possible to unravel the Bloch-Redfield equation into a stochastic Schrödinger equation, the well known secular approximation and the piecewise flat spectral function approximation. This unravelling combines  the connection to the microscopic models of the environment of the Bloch-Redfield equation with the numerical efficiency of the quantum jump unravelling of the Lindblad equation. In the secular approximation the stochastic Bloch-Redfield method can be transformed into a kinetic Monte Carlo simulation when the coupling operators to the bath are given by (or can be approximated by) single transitions between eigenstates of the open quantum system.

We have shown with the example of an SSET that the numerical solutions of the Bloch-Redfield equation and the results of the stochastic Bloch-Redfield method in the secular and piecewise flat spectral-function approximation agree with high accuracy. We also demonstrated the simulation of large systems with our method (over 500 basis states) that are not accessible to numerical solutions of standard master equations.
\begin{acknowledgments}
We thank C. Müller, H. Wiseman and A. Shnirman for useful comments and discussions. N.V. acknowledges the support of the Deutscher Akademischer AustauschDienst (DAAD).
\end{acknowledgments}
\bibliography{/Users/nico/Documents/bibtex/library.bib}

\end{document}